\documentclass[a4paper,12pt]{article}
\usepackage{graphicx,amssymb,bm,latexsym}
\newcommand{\bea}{\begin{eqnarray}}
\newcommand{\ena}{\end{eqnarray}}
\newcommand{\vs}[1]{\vspace{#1 mm}}

\renewcommand{\a}{\alpha}

\renewcommand{\d}{\delta}

\newcommand{\shalf}{\frac{1}{2}}

\newcommand{\nn}{\nonumber\\}
\newcommand{\p}[1]{(\ref{#1})}

\newcommand{\ran}{\rangle}

\begin{document}

\begin{titlepage}

\begin{flushright}
KU-TP 004 \\
hep-th/0607105
\end{flushright}

\vs{10}
\begin{center}
{\Large\bf On the No-ghost Theorem in String Theory}
\vs{15}

{\large
Kazuyuki Furuuchi$^{a,}$\footnote{e-mail address: furuuchi@mri.ernet.in}
and
Nobuyoshi Ohta$^{b,}$\footnote{e-mail address: ohtan@phys.kindai.ac.jp}}
\vs{10}

$^a${\it
Harish-Chandra Research Institute,
Chhatnag Road, Jhusi, Allahabad 211 019, India \\ [.4cm]}
$^b${\it  Department of Physics, Kinki University,
Higashi-Osaka, Osaka 577-8502, Japan\\
[.4cm]}
\end{center}
\vspace{1.5cm}

\centerline{{\bf{Abstract}}}
\vs{5}

We give a simple proof of the no-ghost theorem in the critical bosonic string
theory by using a similarity transformation.

\end{titlepage}
\newpage
\renewcommand{\thefootnote}{\arabic{footnote}}
\setcounter{footnote}{0}
\setcounter{page}{2}

Ten years ago, we proposed the idea of using a similarity transformation to
give a simple proof of the no-ghost theorem in the critical bosonic
string theory~\cite{OF}.
Recently there has appeared a proof of this theorem, which is slightly
different from ours but uses also a similarity transformation~\cite{AK}.
In this short note, we present our proof with some corrections.

The BRST operator of the bosonic string is decomposed in the ghost zero modes
as
\bea
Q_B = c_0 L_0 - b_0 M + d,
\ena
where
\bea
L_0 &=& \a' p^2 + \sum_{n\neq 0} \Big(\frac{1}{2} :\a_{-n}^\mu \a_{n\mu} :
+ n :b_{-n} c_n:\Big)-1, \quad
M = \sum_{n\neq 0} n:c_{-n} c_n :, \nn
d &=& \sum_{n\neq 0} c_{-n} L_n - \frac{1}{2}
 \sum_{\stackrel{m,n \neq 0}{m+n \neq 0}} (m-n):c_{-m}c_{-n} b_{n+m}:\, .
\ena
Here $b_n$ and $c_n$ are the ghost modes with $\{b_n, c_m\}=\d_{n+m,0}$
and $L_n$ is the Virasoro operator for the coordinates:
\bea
L_n = \frac12 \sum_m : \a_m^\mu \a_{n-m,\mu}:\; , \quad
[\a_n^\mu, \a_m^\nu ] = n \eta^{\mu\nu} \d_{n+m,0}.
\ena
The nilpotency of $Q_B$ is equivalent to the relations
\bea
d^2 = M L_0, \quad
[d, L_0] = [d, M] = [M, L_0] = 0.
\label{nilp}
\ena
The physical state is defined by
\bea
Q_B |\mbox{phys}\ran &=& 0,
\label{phys0}
\ena
namely as the cohomology of the nilpotent BRST operator $Q_B$.
The no-ghost theorem claims that the space satisfying this condition
does not involve states of negative norm.

Since $L_0 = \{b_0, Q_B \}$, the physical states obeying the
condition~\p{phys0} also satisfy
\bea
L_0 |\psi \ran = Q_B b_0 |\psi\ran .
\ena
Consequently any physical state is BRST-exact unless it satisfies
the on-shell condition $L_0=0$.
It is convenient to reduce the zero eigenspace of $L_0$ by restricting
to the states annihilated by $b_0$.
In this space, the physical state condition~\p{phys0} reduces to
\bea
L_0 |{\rm phys} \ran = 0, \quad
b_0 |{\rm phys} \ran = 0, \quad
d   |{\rm phys} \ran = 0.
\label{phys}
\ena
Note that we have $d^2=0$ in this space due to the relation~\p{nilp}.

Now we choose our coordinate system such that
$p^i=0 \;(i=1,\cdots, 24)$ and
$p^+\equiv \sqrt{\a'}(p^{25} +p^0) \neq 0$, and define
\bea
\a_n^\pm = \frac{1}{\sqrt{2}}(\a_n^{25} \pm \a_n^0),
\ena
which satisfy $[\a_n^\pm,\a_m^\mp] = n\d_{n+m,0}$.
We also introduce the degree for the oscillators as
\bea
{\rm deg} (\a_n^+,c_n) &=& +1, \nn
{\rm deg} (\a_n^-,b_n) &=& -1,
\ena
and define the degrees for other oscillators and the vacuum to be zero.
The cohomology operator $d$ is decomposed into components with definite
degrees:
\bea
d=d_0+d_1+d_2,
\ena
where
\bea
d_0 &=& p^+ \sum_{n\neq 0} c_{-n}\a_n^-,\nn
d_1 &=& \sum_{n,m,n+m\neq 0} c_{-n}\left[ \a_{-m}^+ \a_{n+m}^-
 + \shalf \a_{-m}^i \a_{n+m}^i  + \shalf (m-n) c_{-m} b_{n+m} \right] ,\nn
d_2 &=& p^- \sum_{n\neq 0} c_{-n}\a_n^+.
\label{op}
\ena
We note that the normal ordering is imposed in the original definition
of the charges but it is not necessary here because all the mode operators
(anti)commute due to the constraints on the sum.
The nilpotency of the operator $d$ gives
\bea
d_0^2 = \{d_0,d_1\} = \{d_0,d_2\}+d_1^2 = \{d_1,d_2\}=d_2^2=0.
\label{nilp1}
\ena

The complication in the no-ghost theorem in string theory comes from the
fact that, in addition to $d_0$ and $d_2$, which are second order in the
mode operators, there are third-order terms in $d_1$.
It was shown that the cohomology of $d_0$ can be extended to that of $d$
by adding terms of higher degrees~\cite{KO}. This procedure was given
perturbatively, and it implies that there is a one-to-one correspondence
between the cohomology of $d_0$ and that of $d$. However, this perturbative
proof is somewhat indirect and cumbersome. The results naturally suggest
that there is a similarity between the cohomologies of these operators.
Here we show that this is indeed the case by explicitly giving
the similarity transformation which maps $d$ into $d_0$,
\bea
e^R d e^{-R} =d_0,
\label{res}
\ena
up to terms trivial in $|$phys$\ran$.
This transformation is useful in the formulation of ``universal string
theory''\cite{univ}, and we expect that it will be useful for other purposes.

After some investigation, we find that
\bea
R &=& \frac{1}{p^+}\sum_{m,n, m+n \neq 0}\Big[
 \frac{m+n}{2nm} \a_{-m}^+ \a_{-n}^+ \a_{m+n}^- \nn
&& +\; \frac{1}{2m} \a_{-m}^+ \a_{-n}^i \a_{m+n}^i
 - \frac{n}{m} \a_{-m}^+ b_{m+n} c_{-n} \Big],
\ena
has the necessary properties~\cite{OF}.
It is then easy to show that
\bea
[R, d_0]=-d_1,\quad
[R, d_1] = 2 d_2 \frac{L_0-p^+p^-}{p^+p^-}, \quad
[R, d_2]=0.
\label{deg2}
\ena
In deriving these results, it must be noted that there are various
terms which drop due to the symmetry of the coefficients, and special
attention must be paid to determining which combinations of the suffices
remain in the sum according to the restriction imposed.
It appears that the second relation is singular for $p^-=0$,
but this is because it is written in terms of the operator $d_2$, which
contains $p^-$, and it is actually a well-defined operator.
The following relations should also be understood in this way.

The result~\p{deg2} means that
\bea
e^R d e^{-R} = d+[R,d] + \frac12 [R,[R,d]] + \cdots
= d_0 + d_2 \frac{L_0}{p^+ p^-},
\label{res1}
\ena
which reduces to Eq.~\p{res} upon using the on-shell condition in Eq.~\p{phys}.
As a consistency check, we note that $R$ commutes with $L_0$ and $M$,
so $e^R d^2 e^{-R} = M L_0$ because of Eq.~\p{nilp}. It is easy to see that
this is true for Eq.~\p{res1} due to Eq.~\p{nilp1} and $\{d_0,d_2\}=p^+ p^- M$.

If we define $K=\frac{1}{p^+} \sum_{n\neq 0}\a_{-n}^+ b_n$,
then $N_0=\{d_0,K\}$ is the level operator for $\a_{-n}^\pm, b_{-n}$
and $c_{-n}$. The states $|\psi\ran$ in the cohomology of $d_0$ satisfy
\bea
N_0|\psi \ran = d_0 K|\psi\ran,
\ena
so that all the states are cohomologically trivial, unless they satisfy
$N_0=0$, or they do not contain these modes.
Thus the cohomology of $d_0$ is spanned by the transverse oscillators
$\a_{-n}^i$ with positive norm, which is denoted by $|{\cal P}\ran$.

According to Eq.~\p{res1}, the physical states of the theory are then given as
\bea
|{\rm phys}\ran = e^{-R}|{\cal P}\ran.
\ena
When we expand the exponent on the right-hand side of this relation, terms
of higher degree appear. Under conjugation, the degree does not change, and
the inner product is nonvanishing only for the case in which the total degree is 0.
This means that the only terms contributing to the inner product are
those of degree 0 made of the transverse oscillators, and hence they give
a positive norm space.
This completes our simplified proof of the no-ghost theorem.

We expect that there would be no essential difficulty in extending our
method to superstrings~\cite{SS}.

\section*{Acknowledgments}

We would like to thank Y. Kazama and Y. Aisaka for pointing out a problem
in the original proof in Ref.~\cite{OF} and for valuable discussions.
The work of N.O. was supported in part by a Grant-in-Aid for Scientific
Research No. 16540250.

\newcommand{\NP}[1]{Nucl.\ Phys.\ {\bf #1}}
\newcommand{\AP}[1]{Ann.\ Phys.\ {\bf #1}}
\newcommand{\PL}[1]{Phys.\ Lett.\ {\bf #1}}
\newcommand{\NC}[1]{Nuovo Cimento {\bf #1}}
\newcommand{\CMP}[1]{Comm.\ Math.\ Phys.\ {\bf #1}}
\newcommand{\PR}[1]{Phys.\ Rev.\ {\bf #1}}
\newcommand{\PRL}[1]{Phys.\ Rev.\ Lett.\ {\bf #1}}
\newcommand{\PTP}[1]{Prog.\ Theor.\ Phys.\ {\bf #1}}
\newcommand{\PTPS}[1]{Prog.\ Theor.\ Phys.\ Suppl.\ {\bf #1}}
\newcommand{\MPL}[1]{Mod.\ Phys.\ Lett.\ {\bf #1}}
\newcommand{\IJMP}[1]{Int.\ Jour.\ Mod.\ Phys.\ {\bf #1}}
\newcommand{\JP}[1]{Jour.\ Phys.\ {\bf #1}}

\vs{5}
Note added:
After this paper was submitted to the arXiv, we were informed that
a discussion of a similarity transformation based on the representations of
Poincar\'e group is given in Ref.~\cite{siegel}.
We thank W. Siegel for pointing this out.

\end{document}